\documentclass[12pt,a4paper]{iopart}

\usepackage{epsfig}
\usepackage{color}
\usepackage{graphicx} 
\usepackage{grffile}
\usepackage{cite}
\usepackage{hyperref}
\usepackage{tikz}
\usepackage{calc}
\newcommand{\cn}{\citeasnoun}
\renewcommand{\cn}{\cite}
\newcommand{\np}{\newpage}

\renewcommand{\Eref}[1] {Eq.\,(\ref{#1})}
\newcommand{\ba}{\begin{eqnarray}}
\newcommand{\ea}{\end{eqnarray}}
\newcommand{\be}{\begin{equation}}
\newcommand{\ee}{\end{equation}}
\renewcommand{\br}{\begin{eqnarray*}}
\newcommand{\er}{\end{eqnarray*}}
\newcommand{\la}{\langle}
  \newcommand{\ra}{\rangle}
\usepackage{bm,amssymb}

\newcommand{\bp}{\begin{minipage}}
\newcommand{\ep}{\end{minipage}}
\newcommand{\hs}{\hspace*}
\newcommand{\vs}{\vspace*}
\renewcommand{\bs}{\bigskip}
\renewcommand{\bs}{\bigskip}
\renewcommand{\ms}{\vs{-5mm}}
\newcommand{\mms}{\vs{-2.5mm}}
\newcommand{\w}{\omega}

\renewcommand{\la}{\langle}
\renewcommand{\ra}{\rangle}

\renewcommand{\k}{\bm k}

\newcommand{\nn}{\nonumber}
\renewcommand{\H}{H$_2$~}
\renewcommand{\contentsname}{\small CONTENTS\mms}
\newcommand{\bt}{\begin{tabular}}
\newcommand{\et}{\end{tabular}}
\renewcommand{\etal}{{\em et al}}
  \newcommand{\mc}{\multicolumn}
\renewcommand{\ns}{\normalsize}

  \newcommand{\ds}{\displaystyle}
   \newcommand{\st}{\small\tt}
   \renewcommand{\t}{\tau}
   \newcommand{\s}{\sigma}
   \newcommand{\q}{\theta}
   
   \newcommand{\g}{\gamma}
   \newcommand{\G}{\Gamma}
   \newcommand{\D}{\Delta}
 \renewcommand{\b}{\beta}
\newcommand{\ig}[1]{\includegraphics[width={#1}]}
\renewcommand{\e}{\epsilon}
\newcommand{\ed}{\end{document}}
\newcommand{\cx}{C$_{60}$~}

\newlength{\ORCIDidheight}
\newlength{\ORCIDidunit}
\definecolor{ORCIDgreen}{HTML}{A6CE39}
\newcommand{\ORCIDid}{%
  \settoheight{\ORCIDidheight}{AXg}%
  \setlength{\ORCIDidunit}{1.2pt * \ratio{\ORCIDidheight}{256 pt}}%
  \raisebox{0.5\depth}{\parbox{\ORCIDidheight}{%
    \begin{tikzpicture}[x=\ORCIDidunit, y=\ORCIDidunit, inner sep=0pt,
        outer sep=0pt]%
      \fill[ORCIDgreen] (128,128) circle (128);
      \fill[white] (70,177) rectangle (86,70);
      \fill[white] (78,200) circle (10);
      \fill[white] (109,177) -- (150,177) %
      .. controls (190,177) and (208,149)%
      .. (208,123)%
      .. controls (208,96) and (186,70)%
      .. (150,70)%
-- (109,70)%
      -- (109,177) -- cycle%
      (124,84)%
      -- (150,84)%
      .. controls (186,84) and (192,110)%
      .. (192,123)%
      .. controls (192,145) and (178,163)%
      .. (150,163) -- (124,163)%
      -- (124,84) -- cycle;
  \end{tikzpicture}%
  }}%
}

\begin{document}

\topical[Resonant photoionization and time delay ]
{Resonant photoionization and time delay}

\author{ Anatoli S.\ Kheifets \,\href{http://orcid.org/0000-0001-8318-9408}{\ORCIDid}}

\address{
	Fundamental and Theoretical Physics, Research School of Physics, \\
	The Australian National University, Canberra 2617, Australia
	}
\ead{a.kheifets@anu.edu.au}
\vspace{10pt}
\begin{indented}
\item[]1 October -- \today
\end{indented}

\begin{abstract}

\ns 

Resonances leave prominent signatures in atomic and molecular
ionization triggered by the absorption of single or multiple
photons. These signatures reveal various aspects of the ionization
process, characterizing both the initial and final states of the
target. Resonant spectral features are typically associated with sharp
variations in the photoionization phase, providing an opportunity for
laser-assisted interferometric techniques to measure this phase and
convert it into a photoemission time delay. This time delay offers a
precise characterization of the timing of the photoemission process.

In this review, a unified approach to resonant photoionization is
presented by examining the analytic properties of ionization amplitude
in the complex photoelectron energy plane. This approach establishes a
connection between the resonant photoemission time delay and the
corresponding photoionization cross-section. Numerical illustrations
of this method include: (i) giant or shape resonances, where the
photoelectron is spatially confined within a potential barrier, (ii)
Fano resonances, where bound states are embedded in the continuum,
(iii) Cooper minima (anti-resonances) arising from kinematic nodes in
the dipole transition matrix elements, and (iv) confinement resonances
in atoms encapsulated within a fullerene cage.

The second part of this review focuses on two-photon resonant
ionization processes, where the photon energies can be tuned to a
resonance in either the intermediate or final state of the atomic
target. Our examples include one- or two-electron discrete excitations
both below and above the ionization threshold. These resonant states
are probed using laser-assisted interferometric
techniques. Additionally, we employ laser-assisted photoemission to
measure the lifetimes of several atomic autoionizing states.

\end{abstract}

\np
\vspace*{-0.5cm}
\footnotesize
\tableofcontents
\ns

\np

\section{Introduction}
\markboth{Introduction}{Introduction}

The studies of resonant photoionization in atoms and molecules have a
long and illustrious history, dating back to the early days of quantum
mechanics. These studies continue to enrich several neighboring fields
of research. Beutler-Fano resonances \cite{Beutler1935,Fano1935} are
now observed in a wide range of physical systems, including
M\"ossbauer nuclei \cite{Li2023}, quantum dots \cite{Kroner2008}, plasmonic
nanostructures \cite{Fan2010,Luk'yanchuk2010,Rahmani2012}, 2D photonic
crystals \cite{Zhou2014}, and metasurfaces \cite{Yang2015}. 
Shape resonances, first discovered by Fermi and Bohr
\cite{Fermi1934,Bohr1936}, are now recognized as a widespread
phenomenon in physics \cite{Rau1968,Connerade1986,Dehmer1988},
chemistry \cite{Langhoff1984}, and biology
\cite{Martin2004dna}. 
Cooper minima \cite{Cooper1964}(anti-resonances), first observed by
Ditchburn \etal \cite{Ditchburn1943} and later explained by Bates,
Massey, and Seaton \cite{Bates1946,Seaton1951}, remain a subject of
intense theoretical and experimental investigation to this day.

\begin{figure}[ht]
\hs{-2.5cm}
\bp{9cm}
\caption{ Spectral and spatial representation of various resonant
  states. A Beutler-Fano resonance is represented as a bound state
  embedded in a continuum (BIC). The shape resonance, characterized by
  a partially confined photoelectron, is depicted as a ``leaky mode.''
  In contrast, a regular bound state is fully confined in space. The
  figure is courtesy of Dr. Kirill Koshelev.
\label{Fig1}
}
\ep
\hs{1cm}
\bp{6cm}
\ig{8cm}{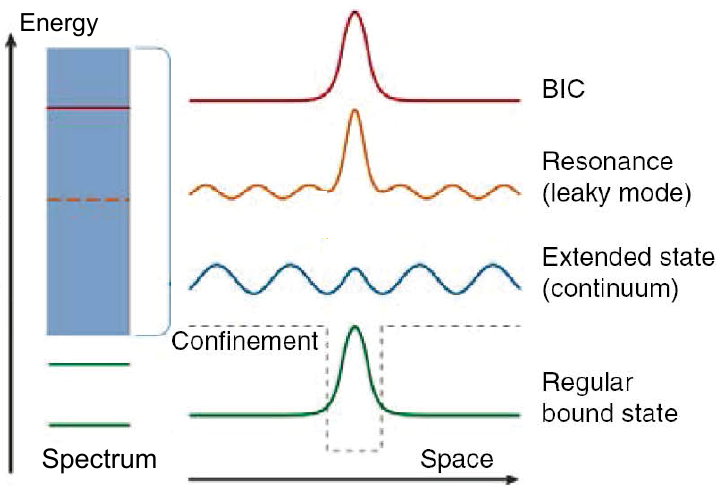}
\ep
\end{figure}

A renewed interest in resonant photoionization has been stimulated by
recent developments in laser-assisted interferometric techniques,
which enable the resolution of atomic and molecular photoionization in
time. One such technique, known as reconstruction of attosecond
beating by interference of two-photon transitions (RABBITT), has
allowed for the measurement of the photoelectron group delay near
shape resonances in various molecules: N$_2$
\cite{haessler09a,Nandi2020,Loriot2021,Borras2023}, N$_2$O
\cite{PhysRevLett.117.093001}, CO$_2$ \cite{PhysRevA.102.023118}, NO
\cite{Gong2022} and CF$_4$ \cite{Ahmadi2022,Heck2021}. A similar shape
resonance measurement in NO \cite{Driver2024} was conducted using an
attosecond angular streaking technique \cite{PhysRevA.106.033106}.
The photoelectron group delay, also known as the Wigner time delay,
was introduced into particle scattering theory
\cite{Eisenbud1948,PhysRev.98.145,PhysRev.118.349} and then extended
to various applications including photoionization (see recent reviews
\cite{deCarvalho200283,Deshmukh2021,Deshmukh2021a,Kheifets2023review}).
In the presence of a resonance, the photoelectron propagation is
naturally delayed relative to the free space propagation. Thus the
Wigner time delay acquires large positive values in the hundred of
attoseconds range (1~as = $10^{-18}$~s).
The RABBITT technique has also allowed for the time resolution of Fano
resonances
\cite{Gruson734,Cirelli2018,Busto2018,Barreau2019,Turconi2020,Neoricic2022}.

A unified approach to resonant photoionization has been offered
recently by considering the analytic properties of the ionization
amplitude in the complex photoelectron energy plane
\cite{Ji2024}. Within this approach, the Wigner time delay can be
directly linked to the corresponding photoionization cross-section, as
was shown earlier in the special case of shape resonances
\cite{PhysRevA.107.L021102}. More generally, this connection can be
made for Fano resonances and Cooper minima \cite{Ji2024}.

In this review article, we revisit \cite{PhysRevA.107.L021102,Ji2024}
and recapitulate the main points of these works. We use a more general
formalism of the complex analysis which illustrates the main results
of \cite{Ji2024} more directly. Our numerical illustrations include
shape resonances in the Xe atom, the I$^-$ ion and the NO molecule,
Fano resonances in the Ne atom, the Cooper minima in Ar and Xe. The
special case of the Ar $3s$ photoionization is particularly
intriguing  as it has attracted a very considerable interest from theory
and experiment. Finally, we consider  confinement resonances in
the Xe atom embedded into the \cx cage.

In the second part, the focus of our review is pointed at
laser-assisted two-photon ionization processes. Here an ionizing XUV
(extreme UV) photon is augmented by an IR laser probe. In such 
two-photon ionization processes, a resonance can occur either in an
intermediate or the final state. The intermediate resonant state can
fall below the ionization threshold in the so-called under-threshold
or uRABBITT process \cite{PhysRevA.103.L011101} or in a
strongly-resonant RABBITT \cite{PhysRevA.104.L021103}. The
intermediate resonant state can also fall into an autoionizing
continuum \cite{han2023laser}. In the most of the RABBITT studies, the
XUV and IR photon energies are tuned to a resonance in the final state
\cite{Gruson734,Cirelli2018,Busto2018,Barreau2019,Turconi2020,Neoricic2022}.
This prfoundly changes the photoelectron spectral and angular
distributions.

While RABBITT has been extensively used for timing characterization of
various resonant ionization processes, it has a limited time span
restricted by the periodic oscillation of its signal. At commonly used
near-IR wavelengths, this time span is generally insufficient to
measure directly lifetimes of most atomic autoionizing states leading
to Fano resonances. Another two-photon XUV/IR ionization process,
known as laser-assisted photoemission (LAPE \cite{Hummert2020}), is
more suited for this purpose. Here we show how LAPE can be used to
determine the lifetime of several most common atomic autoionizing
states \cite{serov2024fano}.

Closely related to the topic, but not covered in the present review,
is the technique of transient absorption spectroscopy. In the present
resonant ionization context, it has been used to provide a universal
phase control \cite{Ott2013,Kotur2016,han2023laser} and to monitor the
birth of a photoelectron near the Fano resonance \cite{Kaldun2016}.

Another technique which is used to time-resolve resonant
photoionization is the attoclock \cite{Tong2022}. However, this
technique is aimed to study the tunneling time and this topic remains
controversial at present
\cite{KheifetsJPB2020,SatyaSainadh2020,Hofmann2021}.

We have to mention other review articles which are related to the
present topic and which should benefit an inquisitive reader. A very
recent article \cite{Ma2024} reviews the concept of attosecond
ionization time delays in strong-field physics.  A recent essay
\cite{Rau2021} connects the energy and time representations in
photoionization and gives a very useful historical and technical
overview. The role of resonant states in many charge-changing
processes in atoms is surveyed in \cite{Lindroth2012}.

Finally, a very extensive literature exists on time resolution of
other resonant ionization processes such as above threshold and
multiphoton ionization \cite{Ge2018,Xu2021,Li2024}, attosecond
streaking \cite{Wickenhauser2005,Goldsmith2018,Borrego-Varillas2021}
and the quantum and coherent control
\cite{Eickhoff2021,Djiokap2021,han2023laser}. The interested reader is
directed to these original articles and references therein.

\np
\section{Single-photon ionization processes}

\subsection{Shape resonances}
\label{Shape}
\markboth{Shape resonances}{Shape resonances} 

\subsubsection{Overview}
\label{overview}

Shape resonances (SRs) have their origins in nuclear physics
\cite{Fermi1934,Bohr1936} , where they are associated with the
collective dipole excitation of a dense nucleon system. Sometimes, SRs
are referred to as giant resonances because they overwhelmingly
dominate the observed cross section over a broad energy range
\cite{Brechignac1994}. In atoms and molecules, the electron density is
insufficient for truly collective excitations, and SRs are instead
associated with the spatial confinement of the photoelectron inside a
potential barrier.

\begin{figure}[ht]
\hs{-2.5cm}
\bp{9cm}
\caption{ Schematic representation of a double-well potential
  associated with formation of SR's.  A potential valley, which is
  degenerate with the ionized continuum, can trap a departing
  photoelectron into a quasi-stationary state. Part of this state is
  leaking out into the continuum \cite{Poccia2011}. The figure is adapted
  from the Research Gate under the Creative Commons Attribution license.
\label{Fig2}
}
\ep
\hs{1cm}
\bp{6cm}
\ig{6.5cm}{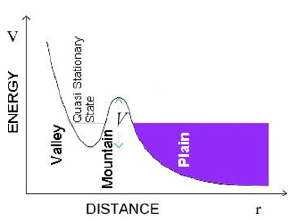}
\ep
\end{figure}

SR's are rather  common  in physics, chemistry and biology
(see Introduction of \cite{Nandi2020} for many diverse
examples). Close to the subject of the present review are SR studies
in electron-molecule scattering \cite{Bardsley1968} and molecular
photoionization \cite{Dehmer1988}. Similar resonant features can be
seen in electron-atom scattering \cite{Shimamura2012} and atomic
photoionization \cite{Rau1968,RevModPhys.40.441,Connerade1986}.
Formation of SR's is well understood
\cite{Bardsley1968,Carlson1983IEEE,Dehmer1988,Child1996,Shimamura2012}.
SR's are associated with the shape of an effective potential in an
open channel which is made of the short-range attractive and
long-range repulsive potentials. Such a double-well potential is
exhibited schematically in \Fref{Fig2}.

A potential barrier $V$ exhibited in \Fref{Fig2} holds a large portion
of the electron wave function while the remaining part of this wave
function leaks out.  Such a combination normally occurs at energies
close to the threshold of an open channel and is typically associated
with a large photoelectron angular momentum $\ell \geq 2$.  Common to
SR's is that they can be turned continuously into bound states by a
slight change of the target Hamiltonian
\cite{Chrysos1998,Horacek2019}. In molecules, SR's are usually
associated with anti-bonding vacant orbitals of the $\sigma^*$
character \cite{Langhoff1984,Piancastelli19991}.

\subsubsection{Electron scattering approach}
\label{scattering}

Attribution of a SR to a particular open channel and its formation by
the photoelectron bouncing off the potential barrier in this channel
offers a convenient representation within the formal electron
scattering theory \cite{PhysRevA.107.L021102}.
The photoionization dipole matrix element $D$ in a single channel
approximation can be expressed via the scattering $T$-matrix. The
latter, in turn, determines the elastic scattering phase. Thus, we can
write
\ba
\label{d-matrix}
D(E)  &=&  d(E) + \int dE' \, d(E') G(E') T(E',E)
\\\nn&\approx& 
d(E) {\rm Im}\, G(E) \  T(E,E)
=
\frac12 \ d(E)
\Big[
e^{2i\delta(E)}-1 
\Big]
\ea
Here we keep the integral term in the right-hand side of
\Eref{d-matrix} and discard the bare term which is negligible near the
resonance. In addition, the Green's function is represented by its
on-shell imaginary part. Our numerical examples, which we will show
below, support both these assumptions.  By squaring the modulus of the
dipole matrix element \eref{d-matrix} we arrive to the cross section
expressed via the scattering phase. The inverse relation allows to
express the scattering phase and the associated Wigner time delay via
the cross-section:
\be
\label{sigma}
\hs{-2cm}
\sigma(E)=\sigma_{\rm max}\sin^2\delta(E)
\ \ , \ \ 
\delta(E) = \sin^{-1} [\sigma(E)/\sigma_{\rm max}]^{1/2}
\ \ , \ \ 
\tau_{\rm W} =\partial \delta(E)/\partial E
\ .
\mms
\ee
Here $\sigma_{\rm max}$ is the cross-section maximum at the resonance
which corresponds to $\delta(k)=\pi/2$.  

\begin{figure}[h]
\hs{-1cm}
\ig{9cm}{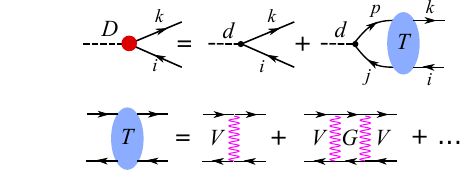}
\mms
\caption{Diagrammatic representation of the integrated dipole matrix
  element $D(E), E=k^2/2$ (top) and the scattering $T$-matrix
  (bottom).  The following graphical symbol are in use: the straight
  line with an arrow to the right and left denotes a photoelectron and
  an ionic (hole) state, respectively. The dotted line exhibits a
  photon, the wavy line stands for the Coulomb interaction. The shaded
  circle and oval are used to represent the $D$- and $T$-matrices,
  respectively. The black dot stands for the bared dipole matrix
  element $D(k)$. The figure is adapted from
  \cite{PhysRevA.107.L021102}.
\label{Fig3}}
\ms
\end{figure}

\begin{figure}[h]
\hs{15mm}
\ig{6.5cm}{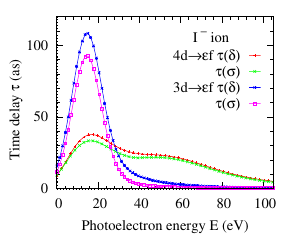}
\hs{5mm}
\ig{6.5cm}{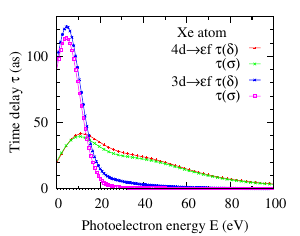}
\mms
\caption{The Wigner time delay in the $3d$ and $4d$ shells of the
  I$^-$ ion (left) and the Xe atom (right). The time delay
  $\t(\delta)$ is expressed as the energy derivative of the
  photoelectron scattering phase in the $nd\to Ef$
  channel. Alternatively, $\t(\sigma)$ is obtained via the
  corresponding cross-section as prescribed by \Eref{sigma}.
  The figure is adapted from \cite{PhysRevA.107.L021102}.
\label{Fig4}}
\ms
\end{figure}

\subsubsection{Numerical examples}
\label{shape_numerical}

For the purpose of numerical illustration, we consider SR's in the
$nd$ shells of the Xe atom and its iso-electronic counterpart, the
I$^-$ ion. In both targets, the departing photoelectron in the
$f$-partial wave is holding strongly by the centrifugal barrier thus
forming the double-well potential exhibited in \Fref{Fig2}. The two
sets of the time delay are shown in \Fref{Fig4}. The one set $\t(\delta)$
is calculated directly using the photoelectron scattering phase in the
given channel. Another set $\t(\sigma)$ is obtained via the
corresponding cross-sections as prescribed by \Eref{sigma}. Both
expressions produce essentially identical results in all the cases.
Details of these calculations can be found in
\cite{PhysRevA.107.L021102}.

\begin{figure}[ht]
\hs{-2.5cm}
\bp{9cm}
\caption{The Wigner time delay $\tau_{\rm W}$ in the NO molecule
  obtained by energy differentiation of the phases derived from the
  corresponding cross-sections. The $\tau(\sigma)$ time delay is
  compared with the Fano formula delays calculated and measured in
  \cite{Holzmeier2021}. The figure is adapted from
  \cite{PhysRevA.107.L021102}.
\label{Fig5}
}
\ep
\hs{1cm}
\bp{8cm}
\ig{7cm}{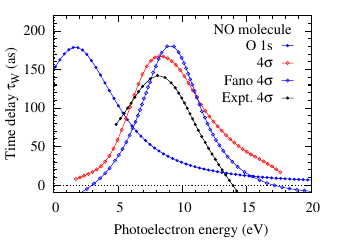}
\ep
\end{figure}


In another numerical illustration, we consider the NO molecule. Here,
the SR occurs because an unoccupied $\sigma^*$ orbital falls into the
$k\sigma$ continuum. A similar $\sigma^*$ resonance is present in the
core shell ionization. To derive the time delay, we use the oxygen
$1s$ \cite{Kosugi1992} and the valence $4\sigma$ \cite{Holzmeier2021}
photoionization cross-sections. In the latter case, we compare the
cross-section derived time delay with the corresponding values which
are calculated and measureed in \cite{Holzmeier2021}.  We observe a
rather close agreement between these three sets. We also note similar
time delays in the valence and core shell photoionization. A very
recent time delay measurement of the oxygen $1s$ shell in NO
\cite{Driver2024} returns considerably larger values which are
strongly dominated by the pure Coulombic time delay. A clear
extraction of the resonant contribution is not possible from this
measurement.

\np

\subsection{Fano resonances}
\label{Fano_resonances}
\markboth{Fano resonances}{Fano resonances} 

\subsubsection{Overview}
\label{fano_overview}

A discrete atomic excitation can fall into an ionization continuum. In
atomic physics, the most familiar examples of this phenomenon are
two-electron excitations in the helium atom and one-electron
excitations in the subvalent shells of heavier noble gases, which
appear above the valence shell thresholds. These ``bound states in the
continuum'' (BIC) manifest as distinct asymmetric lineshapes in atomic
ionization cross-sections and are commonly referred to as Fano
resonances, following the seminal work of Ugo Fano
\cite{Fano1935,PhysRev.124.1866,PhysRev.137.A1364}.

The cross-section near the Fano resonance  takes the form
\be
\sigma(\e) = 
|D(\e)|^2
\propto
{(\e+q)^2\over \e^2+1}
\ \ , \ \ 
\e= {E-E_0\over \Gamma/2}
\ .
\label{Fano}
\ee
Here $\e$ is a detuning from the resonance center $E_0$ measured in
units of the resonance half width and $q$ is the Fano shape
index. The ionization amplitude in \Eref{Fano} can be 
expressed via the phases of the resonant and non-resonant
(background) scattering \cite{Connerade1988,Rau2021}:
\be
D(\e) \propto [e^{2i(\delta+\phi)}-1]/2
\ \ , \ \ 
\cot\delta=\e
\ \ , \ \ 
\cot\phi=q
\ .
\label{De}
\ee
In the absence of the background scattering, $\phi=0$ and the Fano
profile turns into a Lorentzian which is characteristic for an
exponential decay of a discrete excited state with a finite lifetime
$\tau = 1/\Gamma$ \cite{Lorentz1916}.
The Wigner time delay near the Fano resonance is expressed as 
\cite{joachain1975quantum}
\be
\t_W(\e) = {\partial \delta\over \partial E} = {2\over \Gamma}
{1\over \e^2+1} > 0 
\ {\rm irrespective \ of \ } q
\ .
\label{tw}
\ee
Eqs.~\eref{Fano} and \eref{tw} offer a direct link of the time delay
with the cross-section. This link is not so straightforward when a
discrete state is embedded into two or more overlapping continua. This
is the common case of  valence shell ionization of  noble gas
atoms beyond helium. In this case, instead of \Eref{Fano}, the
cross-section is given by a more complex expression
\cite{PhysRev.137.A1364}
\be
\label{Fano1}
\sigma(\e)=\sigma_{0}\left[1-\rho^2+\rho^2
\frac{(q+\e)^{2}}{1+\e^{2}}\right] \ .
\ee
Here $\rho$ is the correlation factor which is required when the
several continuum channels are degenerate at the resonance energy.
\Eref{Fano} is a special case of \Eref{Fano1} with $\rho=1$. There is
no exact analytic expression for the ionization amplitude that would
correspond to the cross-section \eref{Fano1}. An empirical expression
is introduced in \cite{0953-4075-51-6-065008} which agrees reasonably
well with accurate numerical calculations using relativistic
multichannel quantum defect theory.

An alternative approach, which allows the Fano time delay to be
related to the corresponding cross-section in a general case, has been
proposed in \cite{Ji2024}. This approach is based on the analytic
properties of the ionization amplitude in the complex photoelectron
energy plane and is outlined below.

\subsubsection{Kramers-Kronig relation}
\label{kramers}

\begin{figure}[ht]
\bp{7cm}
\ig{6.5cm}{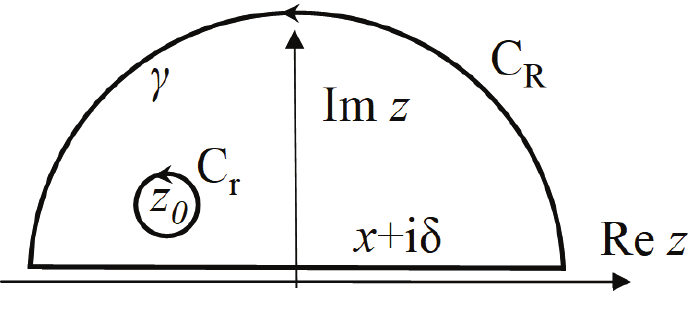}
\ep
\bp{7cm}
\smallskip
\ig{6.5cm}{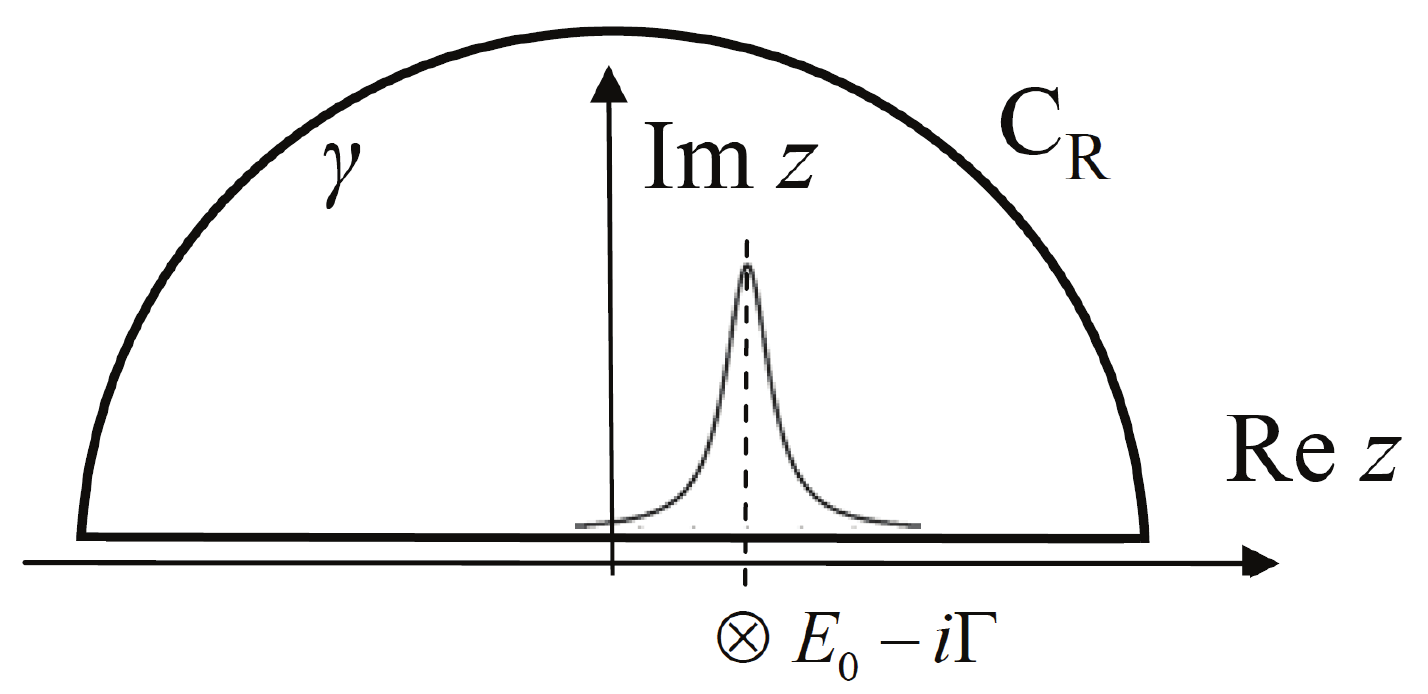}
\ep

\caption{ Schematic representation of the contour $\g$ in the integral
  \eref{Cauchy}. Left: an additional pole is present incide the
  contour. It needs to be isolated by an additional boundary along a
  small circle $C_r$. Right: The resonant cross-section dominates
  strongly over the non-resonant background near the resonance due to
  the pole $E_0-i\Gamma$ below the real axis.
\label{Fig6}
}
\end{figure}

The Cauchy residue theorem equlates the contour integral of an
analytic function $F$ over a boundary $\gamma$ with the sum of the
residues at the poles $a_k$ inside $\gamma$ \cite{watson2012complex}
\be
\label{Cauchy}
 \ds \oint _{\gamma }F(z)\,dz=2\pi i\sum_k {\rm Res} (F,a_{k})
\ .
\ee
We apply the Cauchy theorem to $F(z)=g(z)(z-E+i\delta)^{-1}$ with the
contour $\gamma$ chosen along the upper cut of the real axis and
closed over the half-circle $C_R=Re^{i\phi}$, $\phi\in[0:\pi]$ (see
\Fref{Fig6}).
Then the real axis
integral is vanishing if $g(x)$ is regular inside $\gamma$ and
$g(R)\to0$ rapidly enough as $R\to\infty$. So we can write
\be
\int\limits_{-\infty}^\infty {g(x)dx\over x-E+i\delta}=
\mathcal{P}\!\!\int\limits_{-\infty}^\infty {g(x)dx\over x-E}
-i\pi g(E)
=0
\ee
By separating the real and imaginary parts of this equation, we arrive
to the Kramers-Kronig (KK) relations
\cite{kramers1924law,kronig1926theory,kramers1927diffusion}
\be
\label{KK}
{1\over \pi}\mathcal{P}\!\!\int\limits_{-\infty}^\infty {[{\rm Re/Im}] \
  g(x)dx\over x-E} = \pm[{\rm Im/Re}] \ g(E)
\label{KK}
\ee
The principle value integral in \eref{KK} is also known as the Hilbert
transform (HT) with the Cauchy kernel.

\ms

For the present application, we choose $g(E) = f'(E)/f(E)$, where
$f(E)$ is the photoionization amplitude as a function of the
photoelectron energy $E$.  Then
\be
 g(E) = \frac12{\s'(E)\over\s(E)} + i \tau(E) 
\ .
\ee
If $g(E)$ is regular inside $\g$, then the Hilbert transform returns
the time delay:
\be
\label{LHT}
\tau(E)
= 
\mathcal{H}\left\{\frac12{\s'(E)\over\s(E)}\right\} 
\ .
\ee
However, if $f(z)$ has a pole or node $z_0=a+ib$ inside $\g$, then
$g(z_0)=n(z-z_0)^{-1}$ where $n>0$ for a pole and $n<0$ for a node. To
apply the Cauchy theorem, the pole of $g(z_0)$ needs to be isolated by
integrating over a small circle $C_r$ shown on the left panel of
\Fref{Fig6}. In result, the LHT \eref{LHT} acquires an additional term
\be
\label{LHT1}
\tau(E)
= 
\mathcal{H}\left\{\frac12{\s'(E)\over\s(E)}\right\} 
+ {2\pi n b\over (E-a)^2+b^2}
\ .
\ee
We see that the time delay acquires a Lorentzian component which is
weighted by the displacement of Im$z_0=b>0$ away from the real
axis. 

According to the Cauchy argument principle \cite{watson2012complex},
the sign of $n$, or the net number of nodes $N$ minus the number of
poles $P$, defines the winding number of $f(z)$ equal to the increment
of its phase along the contour $\g$
\be
N-P = {1\over 2\pi i}
\oint_\g {f'(z)\over f(z)} dz = {1\over 2\pi} \Delta_g \arg f(z)
\ .
\ee
In case this number is equal to 0, \Eref{LHT} should be
used. Alternatively, the LHT should be modified as in \Eref{LHT1}.

Applicability of the LHT \eref{LHT} or \eref{LHT1} in a general case
is limited because the photoionization cross-section is only known for
$E\ge0$, whereas the KK relation \eref{KK} requires the knowledge of
$\s(E<0)$.  In a special case of a resonance, exhibited schematically
on the right panel of \Fref{Fig6}, one can neglect a small
non-resonant background away from the pole $E_0-i\Gamma$ below the
real axis. Thus one can close the integration contour over the whole
real axis and the semi-circle $C_R$. This way the KK relation
\eref{KK} applies and the LHT \eref{LHT} or \eref{LHT1} allows to
relate the resonant time delay with the corresponding cross-section.

Results of the preceding Sec.~\ref{Shape} were obtained by utilising
the relation $\sigma_\ell(E) \propto \sin^2\phi_\ell$. This relation
can be derived using a more general formalism of the present section.
In the single-channel scattering case, we can set $f_\ell(E) =
-\cot \left( \phi_\ell(E) \right) $ \cite{LL85}.
Then
\begin{equation}
    \sin^2 \left( \phi_\ell(E) \right) = \frac{1}{{f_\ell(E)}^2+1} = {\Big| \frac{1}{f_\ell(E) \pm {\rm i}} \Big|}^2 ~ .
\end{equation}
Therefore, assuming $f_\ell(E)$ to be analytical,
\begin{eqnarray}
    \mathcal{H} \Bigl\{ \frac{1}{2}\ln \left( \sigma_\ell(E) \right) \Bigr\} = -\arg\{ f_\ell(E)\pm{\rm i} \} 
    = \mp \cot^{-1} \left( f_\ell(E) \right) = \pm \phi_\ell(E)
    \label{eq:Hilbert_sin2_Lorentz}
\end{eqnarray}
If one of the two branches of $(f_\ell(E)\pm{\rm i})$ has no poles or
nodes in the upper half-plane of $E$, we arrive to the desired
relation.  

\subsubsection{Numerical examples}
\label{numerical}

As a numerical illustration, we consider Fano resonances in the $2p$
valence shell of the Ne atom due to discrete $2s^{-1}np$ excitations
from the sub-valent $2s$ shell. These discrete excitations are embedded
into the two degenerate continua $2p\to Es/d$. Correspondingly, the
resonant cross-section is expressed as an incoherent sum of the two
ionization amplitudes
\be
\label{non-coherent}
\s(\e) = \sum_{\ell_f=0,2}
|f_{\ell_f}(\e)|^2
=\s_a+\s_b
{(\e+q)^2\over \e^2+1}
\ .
\ee
In the meantime, the application of the KK relation \eref{KK} and the
LHT \eref{LHT} requires a uniquely defined ionization amplitude which
is not the case for the angular integrated cross-section
\eref{non-coherent}. Such a unique amplitude can be defined in the
case of the angular-resolved ionization process where the
photoionization cross-section is expressed as a coherent
contribution of the two ionization channels.  For example, the
resonant cross-section corresponding to the photoelectron emission in
the polarization direction contains such a coherent sum:
\be
\label{coherent}
4\pi\s(\e,\q=0) = \s(\e)(1+\b)=
\Big|\sum_{\ell_f=\ell_i\pm1}
f_{\ell_f}(\e)\Big|^2=
{A\e^2+B\e+C\over \e^2+1}
\ .
\ee
Here $\b$ is the angular anisotropy  which is expressed near
the resonance via the $A,B,C$ parameters \cite{AK82}.

\begin{figure}[ht]
\bp{\textwidth}
\ig{5.1cm}{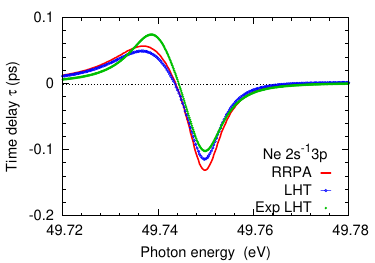}
\ig{5.1cm}{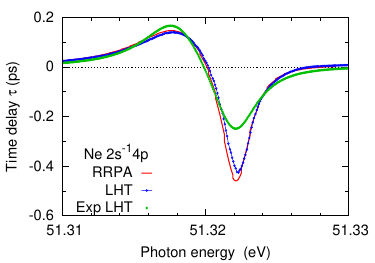}
\ig{5.1cm}{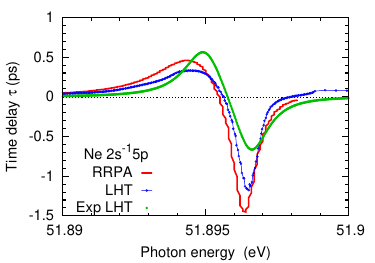}
\ep
\caption{The Wigner time delay of the Ne atom in the region of the
  $2s^{-1}np, n=3,4,5$ (from left to right) resonances corresponding
  to photoelectron emission in the polarization direction. The
  numerical RRPA calculation is compared with the LHT of the RRPA
  cross-section. The LHT with the experimental cross-section
  \cite{Langer1997} is shown for the $n=3$ resonance on the left
  panel.
\label{Fig7}
}
\end{figure}

Our numerical illustration is displayed in \Fref{Fig7}. Here we show
results of the time delay calculations using numerical amplitudes
evaluated in the relativistic random phase approximation
(RRPA). Simultaneously, these amplitudes are used to evaluate the
photoionization cross-section in the polarization direction using
\Eref{coherent}. The latter cross-section is fed into the LHT
\eref{LHT} and converted into the time delay. We see that both sets of
time delay are intimately close to each other. A similar procedure
using \Eref{LHT} and \eref{coherent} is performed with experimental
data from \cite{Langer1997}. A perfect agreement is found between the
present theory and the experiment.

It is notable that, contrary to prescription of \Eref{tw}, the Wigner
time delay in \Fref{Fig6} turns negative. This may seem
counter-intuitive since photoelectron dwelling in a quasi-stationary
resonant state should always delay photoemission but not accelerate
it. However, in the case of the two ionization continua, the
photoemission process becomes more involved. If the two ionization
channels $n\ell\to E\ell\pm1$ are associated with noticeably different
group delays $\partial\delta_{\ell\pm1}/\partial E$, switching from a
``slower'' channel to a ``faster'' one may actually accelerate the
photoemission process. This is indeed the case of the Ne atom where
$\partial\delta_{\ell=0}/\partial E<0$ while $\partial\delta_{\ell=2}/\partial
E>0$. This profound difference of the group delays in Ne and other
noble gas atoms is explained by the Levinson-Seaton theorem (see
\cite{PhysRevLett.105.233002,PhysRevA.87.063404} for more detail).

\np

\subsection{Cooper minima}
\label{Cooper_minima}
\markboth{Cooper minima}{Cooper minima} 

\subsubsection{Overview}
\label{cooper_overview}

Prominent minima in photoionization cross-sections close to the
threshold are observed in many atomic systems (see section 4.5 of
\cite{RevModPhys.40.441} and section 12 of \cite{Starace1983atomic})
These minima are named after John Cooper who established the systematics
of this phenomenon \cite{cooper62a,Cooper1964}. An abnormally small
cross-section near a Cooper minimum (CM) is due to a sign change of the
radial integral in the dipole matrix element.  Such a sign change
usually occurs in the dominant $n\ell \to E\ell+1$ channel
when the bound state $n, \ell+1$ is vacant.

Just by itself, the sign change of the radial integral does not
introduce the phase variation of the photoionization matrix element
exept for a sudden jump by $\pi$. It is due to the coupling of the two
ionization channels $n\ell \to E\ell \pm1$ with their associated
scattering phases $\delta_\ell$ that the net phase of the ionization
amplitude varies strongly near the CM. In the case of a single
ionization channel $ns\to Ep$ like in valence shells of alkali atoms
(see e.g. \cite{Gorczyca2024}), the CM is not associated with a
rapid variation of the phase. However, in a relativistic case, each of the
two spin-orbit split channels $ns_{1/2}\to Ep_{1/2,3/2}$ pass through
their respective CM at a slightly different energy and the
photoionization phase does vary noticeably \cite{Decleva2020}.

The photoemission phase and the Wigner time delay near the CM have
been studied actively in noble gas atoms both experimentally
\cite{PhysRevLett.106.143002,PhysRevA.85.053424,schoun14a,Alexandridi2021}
and theoretically
\cite{PhysRevA.87.063404,DixitPRL2013,PhysRevA.90.053406,PhysRevA.91.063415,Pi2018,Vinbladh2019,Decleva2020}. On
the theoretical side, it is instrumental that the inter-channel
coupling is taken properly into consideration in these studies. One
theoretical approach that provides such a treatment is presented
below.

\begin{figure}[htbp] 
\hs{1cm}
\ig{8cm}{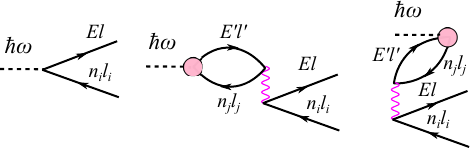}
  \caption{Schematic representation of the dipole
    matrix element $\la El \|\,D\|n_il_i\ra$ in \Eref{amplitude}. The same
    graphical symbols are used as in \Fref{Fig3} except for the shaded
    circle which represents an infinite sequence of diagrams displayed in
    the top row of \Fref{Fig3}. Left: non-correlated dipole matrix
  element.  Center: time-forward process.  Right: time-reverse
  process. Figure adopted from \cite{PhysRevA.87.063404}}.
\label{Fig8}
\end{figure}

\ms\ms
\subsubsection{Random phase approximation with exchange}

The amplitude of the photoelectron emission from the initial atomic
state $n_il_i$ to the final state with a given momentum $\k$ and
energy $E=k^2/2$ is expressed as a coherent sum over various
ionization channels \cite{PhysRevA.87.063404}
\be
\label{amplitude}
f(E,\q)\propto
\sum_{l=l_i\pm1}
e^{i\delta_l}i^{-l}
Y_{lm}(\hat{\k})\,
(-1)^m
\left(\begin{array}{rrr}
l&1&l_i\\
-m&0&m_i\\
\end{array}\right)
 \la El\|D\|n_il_i \ra
\ee
Here the azimuthal angle $\q$ is measured relative to the direction of
the linearly polarized light. 
The dipole matrix element $\la El \|\,D\|n_il_i\ra$
should contain the coupling between various photoemission
channel. Such a coupling is introduced in the random phase
approximation with exchange (RPAE) to the infinite order of the
Coulomb interaction which is exhibited graphically in \Fref{Fig8}.
The phase of the amplitude \Eref{amplitude} is used to evaluate the
time delay
\be 
\label{delay}
\tau(E,\q) = {d\over dE} \arg f(E,\q)\equiv
{\rm Im} \Big[ f'(E,\q)/f(E,\q) \Big] \ .
\ee
This time delay is specific with the photoelectron energy $E$ and the
emission angle $\q$ relative to the polarization direction. 
The same amplitude can be used to evaluate the angular-resolved photoionization 
cross-section
\be
\label{CS}
\sigma(E,\q) = {1\over 4\pi} \sigma(E)
\left[1+\b P_2(\cos\q)\right] \propto |f(E,\q)|^2
\ee
The latter can be plugged to the LHT \eref{LHT} to obtain an
alternative expression for the angular-resolved time delay. 

\ms
\begin{figure}[ht]
\hs{5mm}
\bp{\textwidth}
\hs{-5mm}
\ig{5.5cm}{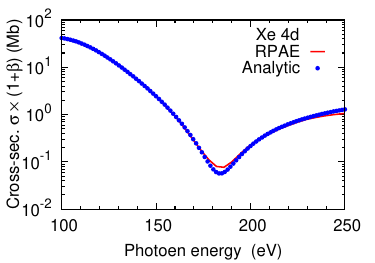}
\hs{-5mm}
\ig{5.5cm}{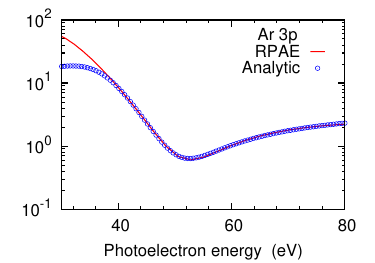}
\hs{-5mm}
\ig{5.5cm}{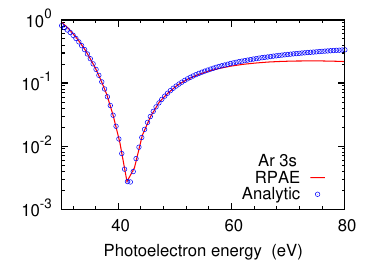}

\hs{-5mm}
\ig{5.5cm}{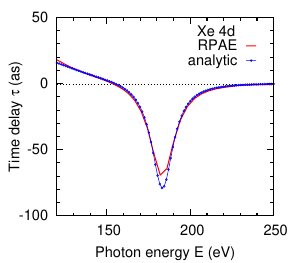}
\hs{-5mm}
\ig{5.5cm}{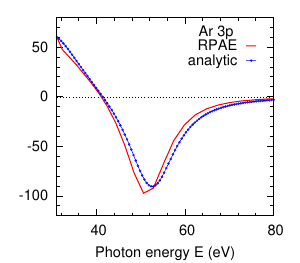}
\hs{-5mm}
\ig{5.5cm}{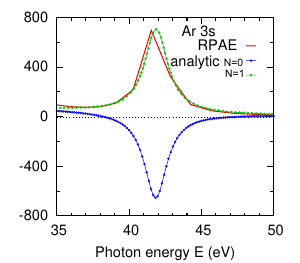}
\ep
\caption{ Comparison of the cross-section-time-delay relation for
  various atomic CM's. The top row of panels display photoionisation
  cross sections in the polarization direction $\sigma(1+\b)$ near the
  CM of Xe $4d$ (left), Ar $3p$ (center), and Ar 3s (right). The RPAE
  calculations (solid red) are fitted with the Fano lineshape
  \eref{non-coherent} (blue dots). The bottom panels display the
  corresponding time delays as calculated by the RPAE (solid red) and
  returned by the LHT (blue dots). In the case of the Ar $3p$ (bottom
  center), the experimental data from \cite{schoun14a} are shown with
  green error bars. In the case of Ar $3s$ (bottom right), the LHT
  with an alternative winding number $N=0$ and $N=1$ is displayed with
  blue and green dots, respectively. The figure is adapted from
  \cite{Ji2024}
\label{Fig9}
}
\end{figure}

\subsubsection{Numerical examples}
\label{cooper_numerical}

We illustrate our technique in \Fref{Fig9} where we display the
photoionization cross-section (top row) and the Wigner time delay
(bottom row) for several atomic CM's: Xe $4d$ (left), Ar $3p$ (center)
and Ar $3s$ (right). Both the time delay \eref{delay} and the
cross-section \eref{CS} are evaluated in the polarization direction
corresponding to $\q=0$. The cross-sections near their respective CM's
are fitted with the Fano lineshape \eref{non-coherent}. The
corresponding Fano parameters are then used to evaluate the time delay
using \Eref{LHT}. The latter is compared with the time delay evaluated
from the RPAE amplitude \eref{amplitude} using the energy derivative
\eref{delay}. In the case of Xe $4d$ (left) and Ar $3p$ (center), the
two sets of the time delay agree exceptionally well. In the latter
case, agreement with experimental data \cite{schoun14a} taken at the two
laser wavelengths of 1.3 and 2.0~$\mu$m is also very good. 

The case of Ar $3s$ is very different. Not only does the corresponding
time delay is nearly an order of magnitude larger than in the two
other cases. The LHT returned time delay is almost an exact sign
inversion of the RPAE calculation. The former is negative as in other
CM's while the latter is positive. This peculiarity of the time delay
in Ar $3s$ CM is related to its distinct nature. While in two other
cases, the phase variation and the time delay arise from the
competition of the two ionization channels $E\ell\pm1$, in the Ar $3s$
case, such a competition is absent. Indeed, there is only one
non-relativistic $Ep$ channel in this case. The Cooper minimium itself
appears due to an inter-channel coupling with $3p\to Es/Ed$ ionization
channels. Such a correlation induced CM is very deep as can be seen in
the top right panel. Correspondingly, the time delay is very large.
As to the sign of the LHT time delay, it depends very strongly on the
winding number of the corresponding ionization amplitude. By changing
this number from zero to one and using \Eref{LHT1} instead of
\Eref{LHT}, the sign of the time delay can be reverted. This way an
agreement with the RPAE calculation is fully restored. In the RPAE,
the Ar $3s$ ionization amplitude has the winding number of 1 as
illustrated graphically in Fig.~5 of \cite{Ji2024}.

\ms
\begin{figure}[ht]
\hs{1cm}
\bp{\textwidth}
\ig{8cm}{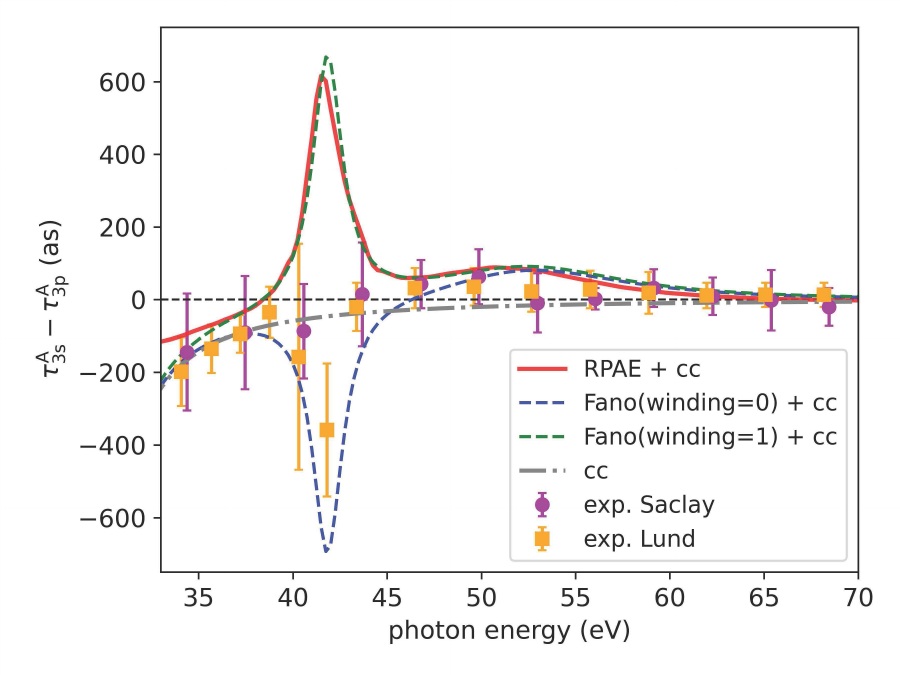}
\ep
\caption{Experimental \cite{Alexandridi2021} and computed (RPAE,
  same as in \Fref{Fig9} relative time delays between argon $3s$ and
  $3p$, compared with the time delay retrieved from the Fano
  parameters. The contribution from the continuum-continuum (CC)
  transition due to the dressing IR field in the RABBIT measurement is
  addressed by adding an additional cc time delay given in
  \cite{0953-4075-47-12-124012}. The figure is adapted from
  \cite{Ji2024}. 
\label{Fig10}
}
\end{figure}

It is instructive to compare various theoretical predictions for the
Ar $3s$ time delay near its CM with the latest experiments. Initially,
both the measurements \cite{PhysRevLett.106.143002,PhysRevA.85.053424}
and several calculations
\cite{PhysRevA.87.063404,PhysRevA.90.053406,PhysRevA.91.063415,Vinbladh2019}
predicted a positive time delay. However, a recent measurement
\cite{Alexandridi2021} hinted at a negative time delay, as shown in
the left panel of \Fref{Fig10}. A very recent measurement by Luo et
al., presented at the International Workshop on Ultra-Fast Science
(Shanghai, 2024, unpublished), strongly indicates a negative time
delay (see the right panel of \Fref{Fig10}). This experimental
observation is supported by several theoretical models. Because the Ar
$3s$ cross-section is very small near the CM, a particular theoretical
model may place the pole inside the integration contour displayed in
\Fref{Fig6}. This instantly changes the winding number of the
ionization amplitude and inverts the sign of the time delay. It
remains to be seen what the final set of Ar $3s$ time delay results
near the CM will reveal in the literature.

A short note is related to the CM of the Na atom as discussed recently
in \cite{Gorczyca2024}. The valence $3s\to Ep$ ionization channel,
which displays the CM, is coupled with inner shell ionization
channels. However, these channels remain closed near the CM and thus
do not alter the phase of the valence shell ionization. Hence the CM
in Na $3s$ is not associated with any noticeable variation of the
resonant phase and the corresponding  time delay. In the complex
photoelectron energy analysis, this corresponds to the pole of the
ionization amplitude appearing on the real axis which invalidates the
application of the LHT.

\begin{figure}[ht]
\bp{\textwidth}
\hs{-5mm}
\ig{5.5cm}{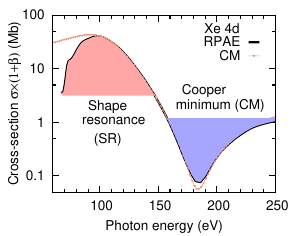}
\hs{-5mm}
\ig{5.5cm}{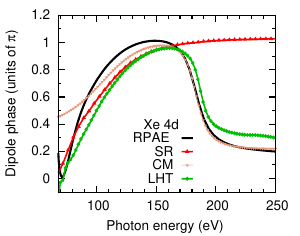}
\hs{-5mm}
\ig{5.5cm}{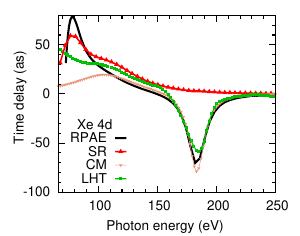}
\ep
\caption{Photoionization cross-section of the Xe $4d$ shell (left),
  the phase of the dipole matrix element (center) and the
  corresponding time delay (right). The shape resonance and the Cooper
  minimum are highlighted in the cross-section plot. Results of a
  multi-channel RPAE calculation are displayed in each channel. This
  calculation is compared with the Fano parameterization of the
  cross-section on the left panel and conversion of the cross-section
  to the dipole phase and time delay in the central and right
  panel. This conversion is performed separately for the SR and CM
  using Eqs.~ \eref{sigma} and  \eref{LHT}. Alternatively, a numerical
  LHT is applied to the cross-section across the whole photoelectron
  energy range. 
\label{Fig11}
}
\end{figure}

\subsubsection{From Cooper minima to shape resonances}
\label{cooper-shape}

In the previous sections \eref{shape_numerical} and
\eref{cooper_numerical}, we related the cross-section with the
corresponding time delay separately for the shape resonances and
Cooper minima. In this short section, we demonstrate the LHT
application across a broad range of the photoelectron energy which
covers both the Cooper minimum and the shape resonance. 
In this application we select $g(E)= \ln f(E)=\ln |f(E) + i\arg f(E)$
in the KK relation \eref{KK}. The LHT returns the photoionization
phase which is then converted to the time delay by the energy
differentiation. In comparison, the LHT \eref{LHT} converts the
logarithmic derivative of the cross-section directly to the time
delay. The former procedure can be more numerically stable than the
latter one.

As the photoionization cross-section and the amplitude are vanishing
at the threshold, the regularization of the LHT is applied following
the recipe of \cite{Burge1974}. The LHT is evaluated numerically using
a Python module $\st Hylbert.py$ from the SciPy library
\cite{2020SciPy-NMeth}. Our numerical results are exhibited in
\Fref{Fig11}. In the left panel, we show the photoionization
cross-section of the Xe $4d$ shell as calculated by RPAE. Correlation
with other ionization channels from the outer $5s$ and $5p$ shells
does not affect the $4d$ cross-section significantly. 
The areas of the SR and CM are shaded for better visibility. The
cross-section at the CM is parameterized using Fano parameters,
although this parameterization loses its accuracy near the
threshold. In the middle panel, we show the phase of the dipole matrix
element. This phase is either calculated directly using the Random
Phase Approximation with Exchange (RPAE) or evaluated by applying the
LHT, either separately over the SR and CM or across the entire
photoelectron energy range. By construction, the separate application
of the LHT is valid only within the SR or the CM. In contrast, the
numerical LHT over the entire photoelectron energy range yields a
phase that closely resembles the RPAE calculation. This phase is
converted to a time delay through energy differentiation, which is
displayed in the right panel. Unfortunately, the energy
differentiation strongly amplifies minor differences in the phase,
resulting in a noticeably different time delay from the RPAE,
especially near the threshold. This example demonstrates the potential
applicability of the LHT across a wide range of photoelectron
energies. However, further improvements to this technique are needed
to achieve quantitatively accurate time delay results.

\np

\subsection{Confinement resonances}
\label{confinement}
\markboth{Confinement resonances}{Confinement resonances} 

\subsubsection{Overview}
\label{confinement_overview}

Confinement resonances (CR's) occur in the photoionization of an
endohedral atom A@\cx incapsulated incide  of a \cx
molecule. This phenomenon was predicted theoretically long ago
\cite{PhysRevA.47.1181}. Since then, it has been studied in depth in
several theoretical works
\cite{0953-4075-38-10-L06,0953-4075-41-16-165001,
  Dolmatov2009,PhysRevA.81.013202}. Recently, CR's have been observed
experimentally in the photoionization of Xe@\cx
\cite{PhysRevLett.105.213001,PhysRevA.88.053402}.

The origin of CR's is well understood. CR's occur due to interferences
between the photoelectron waves emitted directly and those bouncing
off the walls of the encapsulating fullerene \cite{Luberek1996147}.
This multiple scattering shows up prominently as periodic peaks in the
photoionization cross-section \cite{PhysRevA.88.053402}. Similar peaks
are also expected to be present in the corresponding time delay. While
initial investigation of Ar@\cx \cite{PhysRevLett.111.203003} have not
revealed any confinement resonances, subsequent studies on He$^+$@\cx
\cite{Nagele2014}, Xe@\cx \cite{PhysRevA.89.053424,PhysRevA.98.043427}
and Ne@\cx$\!\!^{-q}$ \cite{PhysRevA.94.043401} visualized the CR's in
time delay very clearly.

\begin{figure}[ht]
\hs{2cm}
\ig{7.5cm}{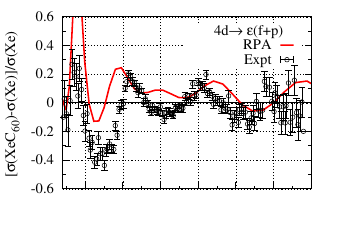}
\ig{7.5cm}{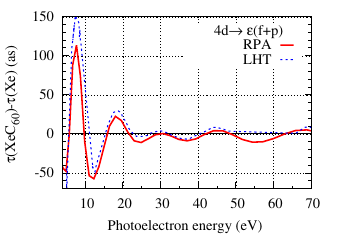}

\caption{Normalized  photoionization cross-section difference
$\rm
\left[
\sigma(Xe@C_{60})-\sigma(Xe)
\right]/\sigma(Xe)
$
(left) and time delay difference $\rm \tau(Xe@C_{60})-\tau(Xe)$
(right) as functions of photoelectron energy. The RPAE calculation is
shown with a solid line. The experimental data in the left panel are
from \cite{PhysRevA.88.053402}. The LHT derived time delay
\cite{Ji2024} is shown with blue dots in the right panel. The figure
is adapted from \cite{PhysRevA.89.053424}.
\label{Fig12}
}
\end{figure}

\subsubsection{Numerical results}
\label{confinement_numerical}

In the Xe@\cx study \cite{PhysRevA.89.053424}, the RPAE approach was
utilized. The effect of the confining \cx on the encaged Xe atom
was approximated by an attractive spherical square well potential.
The numerical result of \cite{PhysRevA.89.053424} are illustrated in
\Fref{Fig12}. In the left panel, the normalized  photoionization
cross-section difference
$\rm
\left[
\sigma(Xe@C_{60})-\sigma(Xe)
\right]/\sigma(Xe)
$
is plotted where it compares favourably with the experiment
\cite{PhysRevA.89.053424}. In the right panel of \Fref{Fig12}, the
resonant part of the time delay $\rm \tau(Xe@C_{60})-\tau(Xe)$ is
visualized. The comparison is made with the LHT result of
\cite{Ji2024}. As was noted in Sec.~\ref{kramers}, the applicability
of the LHT requires that the cross-section vanishes rapidly outside
the resonant region. This requirement is satisfied if we feed into the
LHT the cross-section differene $\rm \sigma(Xe@C_{60})-\sigma(Xe)$
induced solely by the CR. As is seen from \Fref{Fig12}, agreement
between the directly calculated resonant time delay and its
counterpart derived from the LHT of the resonant cross-section is very
good.

\np
\section{Two-photon ionization processes}

\subsection{RABBITT}
\markboth{RABBITT}{RABBITT}
\label{RABBITT}

\subsubsection{Overview}

Two-photon XUV/IR ionization processes offer convenient means to
detect a resonant phase and to convert it to the corresponding time
delay. One such process that had been widely utilized in practice is
reconstruction of attosecond beating by interference of two-photon
transitions (RABBITT). Developed initially for attosecond pulse
characterization \cite{MullerAPB2002,TomaJPB2002}, this technique
found a wide use in time resolution of ionization processes including
the resonant ones
\cite{Gruson734,Cirelli2018,Busto2018,Barreau2019,Turconi2020,Neoricic2022}.

In RABBITT, an ionizing XUV ``pump'' pulse is augmented by a steering
IR ``probe'' pulse. Both pulses are tightly synchronized while their
relative arrival time is varied. In most of RABBITT applications, the
XUV and IR pulses are co-linearly polarized. Recently, RABBITT with
circularly co- and counter-polarized XUV/IR pulses has also been
realized \cite{han2022attosecond,MengHan2024}.

\begin{figure}[ht]

\hs{25mm}
\bp{6cm}
\ig{4.5 cm}{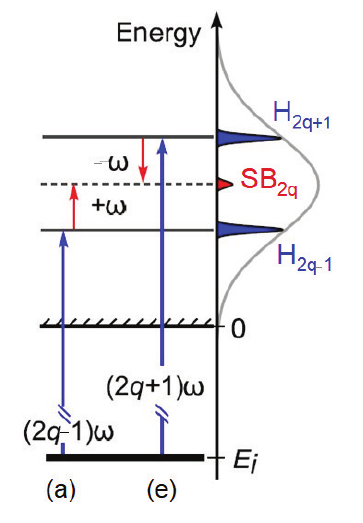}
\ep
\bp{6cm}
\ig{5 cm}{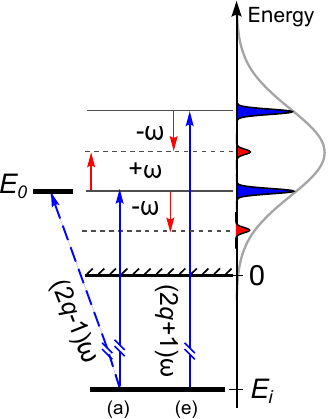}
\bs
\ep

\caption{Left: Photoelectron spectrum of conventional RABBITT is
  formed by absorption of the XUV harmonics H$_{2q-1}$ and
  H$_{2q+1}$. When augmented by an IR photon absorption $+\w$ (a) or
  emission $-\w$ (e) this leads to formation of the sideband SB$_{2q}$
  (adapted from \cn{Kheifets2021Atoms}).
Right: The harmonic H$_{2q-1}$ is tuned to the energy of an
autoionizing state $E_0>0$ embedded in the continuum. This affects the
adjacent sidebands SB$_{2q}$ and SB$_{2q-2}$.
\label{Fig13}
}
\end{figure}

\subsubsection{Conventional RABBITT}

The principle of RABBITT is illustrated schematically in the left
panel of \Fref{Fig13}. In this illustration, a target atom or a
molecule is ionized with a comb of odd XUV harmonics $(2q\pm1)\w$ from
an attosecond pulse train (APT). The spectral harmonic width is
smaller than their separation and the photoelectron spectrum contains
well separated harmonic peaks H$_{2q-1}$ and H$_{2q+1}$. An additional
sideband SB$_{2q}$ is formed in the spectrum once the XUV photon
absorption is augmented by absorption (a) or emission (e) of a single
photon $\pm\w$ from the driving IR pulse.  As the two (a/e) interfering
quantum paths lead to formation of the same SB, its height
oscillates as the XUV/IR relative arrival time $\Delta$ varies:
\be
S_{\rm SB}(\Delta) =A+
B\cos[2\omega\Delta-C]
\ \ , \ \
C = 2\w\t_a
\ \ , \ \
\t_a = \t_W+\t_{cc}
\ .
\label{RABBITT}
\ee
Here  $A,B$ are the RABBIT magnitude parameters whereas $C$ is the
RABBITT phase. The latter can be converted to the atomic time delay
$\tau_a$ which is composed of the Wigner time delay and the
continuum-continuum (CC) correction \cite{Dahlstrom2012}.

Conventional RABBITT can become resonant if one of the harmonic peaks
H$_{2q\pm1}$ overlaps with a bound state embedded in the continuum. In
the right panel of \Fref{Fig13} the H$_{2q-1}$ harmonic becomes
resonant with the energy of an autoionizing state $E_0>0$. Such a
resonant RABBIT scheme has been realized in several experimental
studies performed in the near-IR spectral range (wavelength of 800~nm).
In one such study \cite{Gruson734}, H$_{35}$ was tuned to the $2s2p$
autoionizing state of He ($sp2+$ as classified in
\cite{cooper1963classification}) Similarly, by tuning appropriately
the IR carrier frequency, both the $sp2+$ and $sp3+$ autoionizing
states of He could be probed \cite{Busto2018}.
In heavier noble gases, H$_{17}$ was tuned to $3s^{-1}4p$ autoionizing
state of Ar \cite{Kotur2016}. By a fine adjustment of the IR photon
frequency, the whole sequence of $3s^{-1}np$ resonances of Ar could be
probed \cite{Cirelli2018}. With a sufficient energy resolution, such a
measurement was able to resolve spin-orbit splitting
\cite{Turconi2020}. In a similar measurement \cite{Barreau2019}, the
autoinizing $2s^{-1}np$ states of Ne were probed.

\begin{figure}[ht]

\bp{1.1\textwidth}
\ig{5.5 cm}{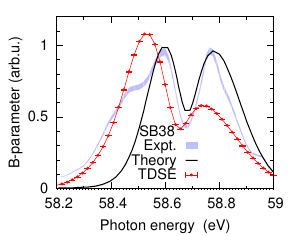}
\hs{-5mm}
\ig{5.5 cm}{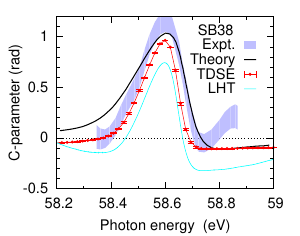}
\hs{-5mm}
\ig{5.5 cm}{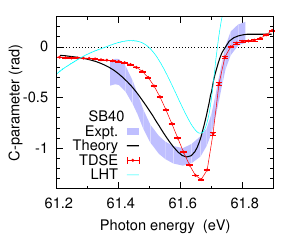}
\ep

\caption{Resonant $B$ (left) and $C$ (center and right) parameters in
  He near the $sp2+$ resonance. The experiment and theory from
  \cite{Busto2018} are compared with a multi-channel solution of the
  TDSE from \cite{serov2024fano}. The
  $C$ parameters from the numerical LHT calculations are displayed in
  the middle and right panels. 
\label{Fig14}
}
\end{figure}

Appearance of a Fano resonance in one of the arms of an XUV/IR
interferometric process change the RABBITT magnitude and phase
parameters in a very profound way. A general theoretical approach
to resonant RABBITT is outlined in
\cite{PhysRevLett.113.263001,PhysRevA.93.023429,Argenti2017}.  In
\Fref{Fig14}, accurate numerical simulations illustrate resonant
modification of the RABBITT $B$ and $C$ parameters near the $sp2+$
resonance of He.  Experiment and theory from \cite{Busto2018} are
compared with a multi-channel solution of the time dependent
Schr\"odinger equation (TDSE) from \cite{serov2024fano}. The
wavelength of the IR probe is tuned in such a way that H$_{39}$ is
resonant with $sp2+$ which affects the adjacent SB$_{38}$ and
SB$_{40}$ as shown in the Figure.  Away from the resonance, the
resonant and non-resonant SB behave very similarly. This can be used
as in the case of confinement resonances considered in
Sec.~\ref{confinement_numerical}. The resonant/non-resonant difference
signal is fed into a numerical LHT to get a net resonant contribution
to the RABBITT phase (the $C$ parameter). This contribution is shown
in the middle and right panels of \Fref{Fig14} where it qualitatively
agree with accurate numerical calculations.

\begin{figure}[ht]

\hs{25mm}
\ig{4 cm}{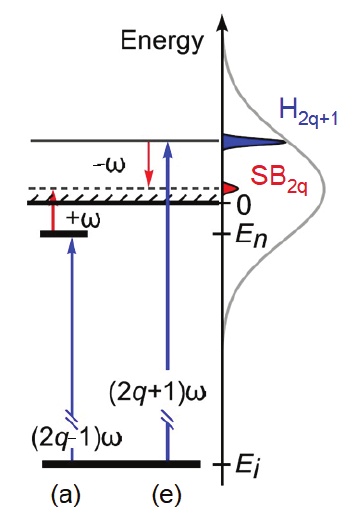}
\ig{13cm}{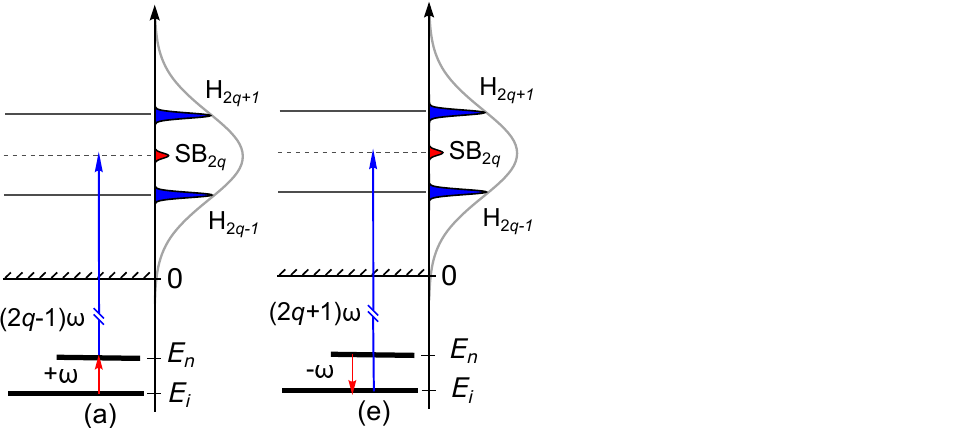}

\caption{ Left: Under-threshold RABBITT proceeds by absorption of an
  IR photon from a bound state $E_n<0$ to the sideband
  SB$_{2q}$ (adapted from \cn{PhysRevA.103.L011101}).
Center: Strongly resonant RABBITT proceeds by absorption of
an IR photon $\w$ from the ground $E_i$ to the resonant $E_n$ state
Right: Same process is facilitated by emission of an IR photon from
the resonant $E_n$ state to the ground state $E_i$.
(adapted from \cn{PhysRevA.104.L021103})
\label{Fig15}
}
\end{figure}

\subsubsection{Under-threshold RABBITT}
\label{uRABBITT}

If one harmonic energy submerges below the ionization threshold
$(2q-1)\w <|E_i|<(2q+1)\w
\,,
$ 
the corresponding harmonic peak H$_{2q-1}$ disappears from the photoelectron
spectrum. Instead, the missing absorption path of the
conventional RABBITT process can proceed via a discrete atomic excitation
$E_n<0$. Such an under-threshold or uRABBITT process is illustrated
graphically in the left panel of \Fref{Fig15}.

\begin{figure}[htb]
\hs{1.5cm}
\ig{6 cm}{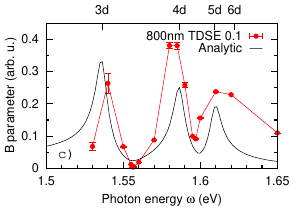}
\hs{-0.5cm}
\ig{6 cm}{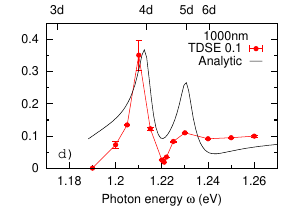}

\hs{1.5cm}
\ig{6 cm}{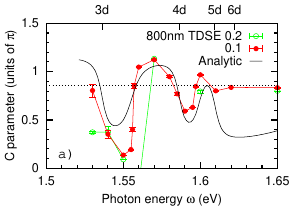}
\hs{-0.5cm}
\ig{6 cm}{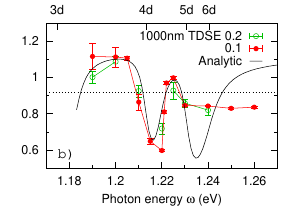}

\caption{The RABBITT magnitude $B$ (top) and $C$ (bottom) parameters
  in Ne as the functions of the fundamental laser frequency in the
  800~nm (left) and 1000~nm (right) wavelength ranges.  The top
  horizontal axis marks the crossing of the harmonic peaks H13 (left)
  and H17 (right) with the discrete $nd$ energy levels. Numerical TDSE
  simulations with various spectral width of the APT (in eV) are
  compared with a simple analytic model. The figures are adapted from
  \cite{Kheifets2021Atoms}.
\label{Fig16}}
\end{figure}

The uRABBITT can be observed experimentally in He
\cite{SwobodaPRL2010,Neoricic2022} where H$_{15}$ goes under the
threshold whereas H$_{17}$ is used as a complementary interferometric
arm. This way a sequence of discrete $1snp \ ^1P_1$ excitations with
$n=3,4,5$ can be reached. 
An analogous approach can be utilized in Ne where H$_{13}$ goes under
the threshold and H$_{15}$ is used as its over-the-threshold partner
\cite{Villeneuve2017,moioli2024role}. In \cite{Villeneuve2017}, the
population of various $2p^{-1}3d_m$ discrete sub-states can be
monitored.
Very similarly, H$_{5}$ and H$_{7}$ can be used to realize a uRABBITT
in the Ar atom \cite{Kheifets2023}. 

The bound state structure of the target can be deduced from an
under-threshold $u$RABBITT process
\cite{PhysRevA.103.L011101,Kheifets2021Atoms}. Such a determination is
illustrated in \Fref{Fig16} where the magnitude $B$ and phase $C$
parameters of Ne are displayed in the region of $2p^{-1}nd$ excitations.
These excitations reveal themselves as sharp peaks of the magnitude of
the RABBITT oscillations and cause a sharp variation of the RABBITT
phase.

\subsubsection{Strongly resonant RABBITT}
\label{LiRABBITT}

The crossing of the submerged harmonic H$_{2q-1}$ with a bound state
in the uRABBITT process affects profoundly just a single SB$_{2q}$. A
strong modification of the whole RABBITT spectrum can be achieved in
the strongly-resonant RABBITT process displayed in the middle and
right panels of \Fref{Fig15}. Here a discrete excitation is resonant
with the IR carrier frequency $|E_n-E_i|=\w$. In a strongly-resonant
RABBITT process, the two-arm interference is realized by the XUV
absorption from the ground state $(2q+1)\w$ or the excited state
$(2q-1)\w$. As such, it affects all the SB$_{2q}$ for $\forall q$.  In
addition, it does not contain a CC transition which is always present
in the conventional or an under-threshold RABBITT. Instead of the CC
component, the strongly resonant RABBITT phase contains the resonant
contribution  which can be approximated as \cite{PhysRevA.103.L011101}
\be
\label{Eq5}
\arg \Big[ \w+E_{i}-E_{n}-i\Gamma\Big]^{-1}
=\arctan(\Gamma/\Delta)
\ .
\ee
Here $\Gamma$ is the spectral width of the IR pulse and $\Delta
\equiv\w+E_{i}-E_{n}$ is the detuning.

\begin{figure}[ht]

\bp{8cm}
\caption{ The $C_{8}$ phase variation with the photon energy $\w$ is
  plotted for $2p_{m=0,1}$ and $2s$ initial states. The dotted line
  visualizes \Eref{Eq5}. The figure is adapted from
  \cn{PhysRevA.104.L021103}.
\label{Fig17}
}
\ep
\hs{5mm}
\bp{7cm}
\ig{7cm}{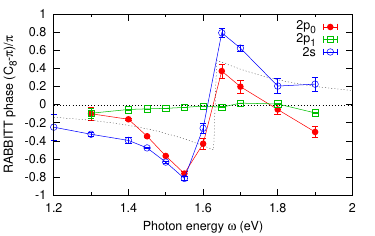}
\ep

\end{figure}

A strongly resonant RABBITT process can be observed in the Li atom
where the $2s^{-1}2p$ excitation is resonant in the 800~nm wavelength
range \cite{PhysRevA.104.L021103,PhysRevA.105.063110}.  An
illustration of such a process is displayed in \Fref{Fig17} where the
RABBITT phase (the $C$ parameter) is plotted for the SB$_8$. The Li
atom is prepared initially either in the ground $2s$ or an excited $2p_m,
m=0,1$ states. For the $m=0$ initial states, the $2s-2p$ resonance
affects very strongly the RABBITT phase in a narrow energy range. 
For the $2p_1$ initial state, such a resonance is absent and the
corresponding RABBITT phase remains flat.

\np

\subsection{Laser-assisted photoemission}
\markboth{LAPE}{LAPE}
\label{LAPE}

\subsubsection{Overview}

RABBITT oscillation \eref{RABBITT} repeats itself periodically. So the
useful time span of the RABBITT signal is $\pi/\w\simeq1.3$~fs at the
800~nm laser wavelength.   Meanwhile, the lifetime $\t$ of most 
atomic autoionizing states is about an order of magnitude larger as
shown in \Tref{T1}. So a direct application of the RABBITT process for
determination of $\t$ is of little use.

\begin{table}[h]
\bp{8cm}
\caption{The width $\Gamma$ (in meV) and the lifetime $\t$ (in fs) of
  various molecular and atomic autoionizing states (AIS). The
  literature $\t$ values are compared with the LAPE determination \cite{serov2024fano}. 
\label{T1}
}
\ep
\hs{5mm}
\bp{8cm}
\small
\bt{lrrrcc}
\mc{2}{c}{AIS} &Width&\mc{2}{c}{Lifetime}\\
&&$\Gamma$, meV &\mc{2}{c}{$\t$, fs}\\
&&&Ref. & LAPE\\
\hline\hline\\
H$_2$ & $^1\Sigma_g+$ & 971 & 0.7\cite{Bottcher1974}\\
Li$^+$& $2s2p+$ &74 & 8.7\cite{Carroll1977} & 9.2\\
He& $2s2p+$ & 37 & 17\cite{Domke1996} & 15\\
He& $2s3p+$ &8.4&82\cite{Domke1996} & 80\\
\et\ns
\ep
\end{table}

An alternative technique of laser-assisted photoemssion
(LAPE) \footnote{The acronym LAPE can be found in preceding
  literature, see e.g. \cite{Hummert2020}} can be used for this
purpose. In LAPE, in comparison to RABBITT, the APT is replaced with
an isolated XUV pulse. Instead of a sequence of SB$_{2q}$, only one
pair SB$_{\pm1}$ is formed which corresponds to absorption $+\w$ or
emission $-\w$ of a single IR photon.  Such a technique has been used to
determine the lifetime of the Auger decay of the Kr atom
\cite{Drescher2002}. Unlike the Fano process, this decay is purely
exponential and determination of its time constant is straightforward.
An extension of this technique to AIS was proposed in \cite{Zhao2005}.
The time representation of the Fano amplitude \eref{De} is given by the
expression
\be
F(t)=\frac{\Gamma}{2}(q-i) e^{-i E_0 t-\Gamma t/ 2}+i \delta(t-0)
\label{Ft}
\ee
Here the first term in the right hand side describes an exponential
decay of an AIS whereas  the second term is responsible for an
instantaneous photoemission. In the absence of the latter process,
\Eref{Ft} describes exponential decay of a discrete excited state with
a finite lifetime $\tau = 1/\Gamma$ \cite{Lorentz1916}.
\Eref{Ft} is restricted to the special case of \Eref{Fano} in which an AIS is
embedded into a single ionizing continuum. A more general case of
several  continua degenerate with an AIS is considered below.

\subsubsection{LOPT formalism}\hs{-3mm}.

In this section, we present the formalism within the lowest order
perturbation theory (LOPT) as outlined in \cite{serov2024fano}. In
brief, the time-dependent LAPE amplitude can be expressed as
\be
a_f(t) = (-i)^2 \sum_{n\ne i} 
\int_{-\infty}^{t} \hs{-3mm}dt'V_{fn}(t') e^{i(E_f-E_n)t'} 
\int_{-\infty}^{t'}\hs{-3mm} dt''  V_{ni}(t'') e^{i(E_n-E_i)t''} 
\ .
\label{aft}
\ee
The perturbation matrix elements 
$V_{ab}(t)$
contain  the dipole interaction with the   XUV and
IR fields 

We tune the carrier frequency of the XUV pulse to the excitation
energy of the AIS $\omega_x \approx E_0 - E_i$. We also assume that
the duration of the XUV pulse is much smaller that the lifetime of the
AIS $T_x\ll \t$ whereas the duration of the IR pulse $T\sim \t$. In
this case, the latter state can be considered as stationary during its
interaction with XUV pulse. Under these assumptions, the LAPE
amplitude \eref{aft} can be transformed to the following expression:
\ba \nn a_{\pm1}(\e) &\sim& \exp \left[-
  \frac{T^2(2i\e\tau-1)^2}{8\tau^2} - i\e \D - \frac{\D}{2\tau}\right]
\\ &\times& 
\left[\mathrm{erf}\left\{\frac{T(2i\e\tau-1)}
  {\sqrt8\tau}+\frac{\D}{\sqrt2T}\right\}+1\right] 
\ .
\label{gauss}
\ea
The amplitude \eref{gauss} describes the population of the SB$_{\pm1}$
with a detuning $\e=E_f-E_0 \pm \omega$ and corresponding to the
XUV/IR delay $\D$ \footnote{In \Eref{RABBITT} $\D$ denotes the carrier
  frequency delay which is physically different from the envelope
  delay in \Eref{gauss}}.
At $\D \gg T$  the SB acquires a  Gaussian lineshape
\be A_{\rm SB}(\e)=|a_{\pm1}|^2 \sim \exp\left[- T^2\e^2 -
  \frac{\D}{\tau}\right] 
\ .
\label{gauss1}
\ee
The width of the Gaussian is determined by the length of the IR pulse
$T$ but not the width of the resonance $\Gamma$. The magnitude of the
Gaussian decreases exponentially with increasing $\D$. The time
constant of this exponential decay is equal to the lifetime of the
autoionizing state $\tau$.

\begin{figure}[htb]
\bp{1.1\textwidth}
\hs{-5mm}
\ig{6cm}{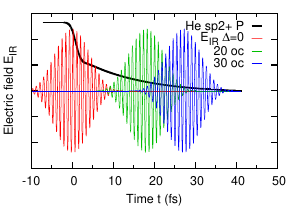}
\hs{-5mm}
\ig{6cm}{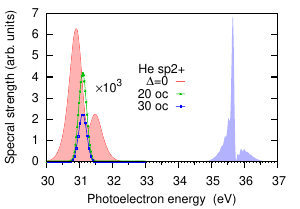}
\hs{-5mm}
\ig{6cm}{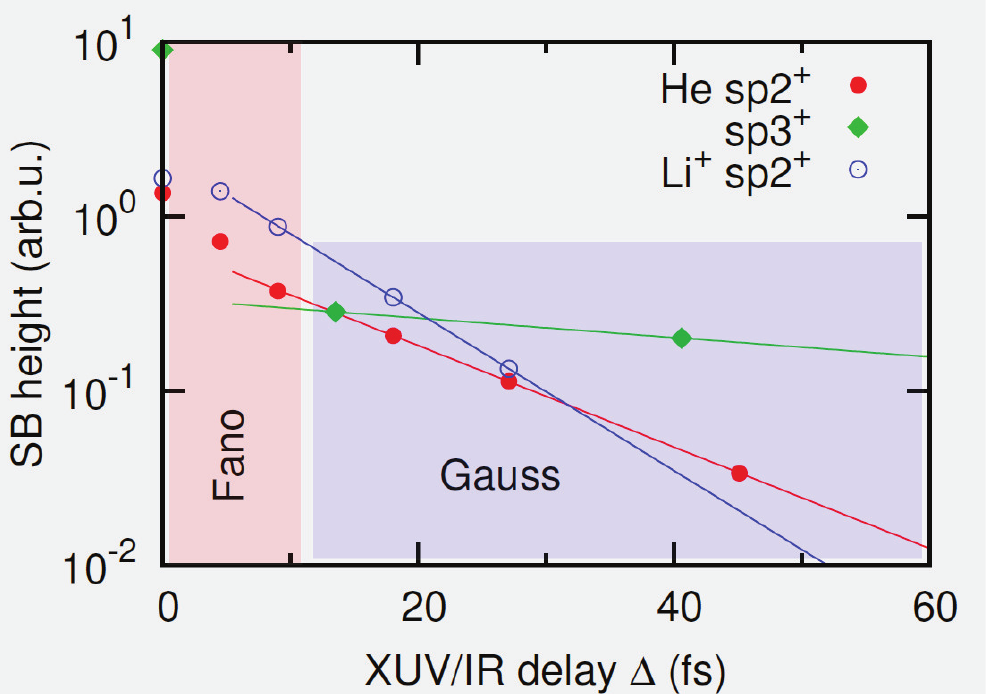}
\ep

\caption{Left: the thick solid line shows the scaled probability $P$ of
  locating the departing photoelectron within the simulation
  boundary. Differently colored IR probes arrive at various delays
  $\D$ (in units of optical cycles, $\rm 1~oc = 0.9~fs$). Middle: the
  photoelectron spectrum corresponding to different XUV/IR delays
  $\D$. The SBs (only one set is shown) are scaled by a factor $10^3$
  for better visibility. Right: the SB height as a function of the
  XUV/IR delay is fitted with an exponential decay function. The
  shaded areas mark the Fano and Gauss lineshape appearance. The top
  and middle panels display the He $sp2^+$ data whereas the bottom
  panel shows the He $sp2^+, sp3^+$ and Li$^+$ $sp2^+$ data. The
  figures are adapted from \cite{serov2024fano}.
\label{Fig18}}
\end{figure}   

Qualitatively, the lineshape \eref{gauss1} is easy to understand.  In
the Fano resonant ionization process, a decaying AIS is embedded in
the continuum resulting in a characteristic asymmetric photoelectron
lineshape. In LAPE, the resonant part of the Fano process is
transferred to the adjacent SB's. At sufficiently long XUV/IR delay
$\D\to\infty$, the IR pulse will not alter the energy of the
photoelectron. Classically, a free electron, far away from the
nucleus, cannot absorb a photon to conserve both the momentum and
energy. In result, resonant part of the SB's becomes decoupled from
the embedding continuum.  Hence, the sideband lineshape becomes
symmetric and its magnitude decreases exponentially with the timing
constant of the AIS decay.

\subsubsection{Numerical illustration}

This metamorphosis of the SB lineshape and the exponential decay of
its heifht is illustrated graphically in \Fref{Fig18}. In the left
panel of the figure, we display the scaled probability of locating the
departing photoelectron within the simulation boundary.  This
probability decreases at time after arrival of the ionizing XUV pulse
at $t=0$ .  Differently colored IR probes arrive at various delays
$\D$ (in units of optical cycles, 1~oc=0.9~fs at the presently used
wavelength of 266~nm). The photoelectron spectrum is displayed in the
middle panel of \Fref{Fig18}. The main photoelectron line (shaded
purple) is insensitive to the XUV/IR delay $\Delta$. In the meantime,
SB$_{-1}$ (scaled for better visibility) changes both its shape and
height. The shape changes from an asymmetric Fano-like to a simple
Gaussian and the height decreases steadily with $\G$. This decrease is
illustrated in the right panel for several AIS. At small $\D$, where
the lineshape is Fano-like, there is no clear systematic in the height
measurement. However, when the SB acquires a symmetric Gaussian
lineshape, the decay becomes clearly exponential and the lifetime
constant can be easily deduced by a simple fit $\exp(-t/\t)$. Such a
determination returns the $\t$ results tabulated in \Tref{T1}  which are
very close to the literature values.

In principle, the  TDSE code used for the present
simulations \cite{Serov2024} utilizes a molecular multi-configuration
expansion. Hence, it can describe the AIS of the \H molecule. However, its
lifetime is too short to fit it with an exponential decay formula
under the present simulation conditions.

\np
\section{Summary and outlook}
\label{Summary}

This review article briefly explores various aspects of resonant
ionization processes induced by single- and two-photon absorption. In
single-photon ionization, several types of resonances are considered,
including shape or giant resonances, Fano resonances, Cooper minima
(which can be viewed as anti-resonances), and confinement
resonances. These resonant ionization processes are unified through an
analytic approach based on the Kramers-Kronig relation, enabling the
application of the logarithmic Hilbert transform. This approach allows
for relating the photoionization cross-section to the corresponding
time delay, effectively ``converting megabarns to attoseconds.''
Such a conversion connects traditional ``old'' photoionization studies
performed using synchrotron sources with the ``new'' attosecond
physics driven by laser-assisted interferometric
techniques. Techniques such as RABBITT and LAPE are discussed in
detail. While RABBITT enables the derivation of the resonant phase,
LAPE is instrumental in the accurate determination of the lifetimes of
various autoionizing states.

In the presented applications, the resonances were either isolated or
subtracted from the non-resonant background. Attempts to apply the
logarithmic Hilbert transform across a wider range of photoelectron
energies have had mixed success, and the numerical technique requires
further development to fully exploit this approach. Additionally, the
case of several overlapping resonances, a characteristic feature of
molecular ionization, needs to be addressed in future applications.

Time-resolved photoemission studies encompass a wide range of topics,
and this review touches on only a few selected aspects of this rapidly
expanding field. Nevertheless, the author hopes that the unified
approach presented here will be helpful and stimulate further interest
and applications within the broader atomic physics community.

\section*{Acknowledgment} 
\markboth{Acknowledgment}{Acknowledgment}

The idea for this review article was conceived during the author's
presentation at the Seminar on Theoretical Chemistry, Molecular
Spectroscopy, and Dynamics at ETH Z\"urich. The author is very grateful
to the host, Prof. Hans Jakob Wörner, for his warm hospitality. The
author also acknowledges other members of the ETHZ group,
Prof. Kiyoshi Ueda, Dr. JiaBao Ji and Dr. Meng Han for many
stimulating discussions.

Multiple interactions with several participants at the International
Workshop on Photoionization (IWP-24) in Ascona, Switzerland, were
highly beneficial. The author wish to thank Prof. Thomas Gorczyca (Western
Michigan University), Prof. Dajun Ding (Jilin University), Dr. David
Busto (Lund University) and Dr. Taran Driver (Stanford University).

The author acknowledges former members of his group who significantly
contributed to the theoretical results reviewed in this article:
Dr. Igor Ivanov, currently at the Institute for Basic Science in
Gwangju, South Korea, and Dr. Vladislav Serov, now at Saratov State
University, Russia. The author also benefited from collaborations with
other theoretical groups led by Prof. Steve Manson at Georgia State
University and Prof. Pranawa Deshmukh at the Indian Institute of
Technology, Tirupati.

Finally, the author and his group extensively relied on the
computational resources of the National Computational Infrastructure
facility.

\np

\section{References}
\markboth{References}{References}

\providecommand{\noopsort}[1]{}\providecommand{\singleletter}[1]{#1}%
\providecommand{\newblock}{}

\end{document}